\documentstyle [12pt]{article}
\date\today

\def\vld{Vl\'{a}dar}
\def\noz{Nozi\`{e}res}

\def\ufp{U$^{4+}$}
\def\ctp{Ce$^{3+}$}
\def\ucon{$5f^2(J=4)$}
\def\ccon{$4f^1(J=5/2)$}

\def\gse{\Gamma_7}
\def\gei{\Gamma_8}
\def\gni{\Gamma_9}

\def\gth{\Gamma_3}

\def\gfi{\Gamma_5}
\def\gsi{\Gamma_6}

\begin{document}

\parindent=0in
{\Large{\bf  Selection Rules for Two-Channel Kondo Models of U$^{4+}$ and
Ce$^{3+}$ Ions in Metals}}\\

D.L. Cox, Department of Physics, 174 W. 18th Avenue,\\
The Ohio  State University, Columbus, Ohio, USA, 43210\\

Symmetry based selection rules are developed providing minimal criteria for the
existence of two-channel Kondo interactions between conduction
electrons and the low
energy degrees of freedom on U$^{4+}$ and Ce$^{3+}$ in a metal host, assuming
that the underlying microscopics are regulated by the Anderson
Hamiltonian.  An additional dynamic selection rule is imposed on Ce$^{3+}$
ions.
The selection rules restrict the two-channel quadrupolar Kondo effect
to U$^{4+}$ ions in cubic, tetragonal, and hexagonal symmetry.  For hexagonal
and tetragonal symmetry, the Kondo effect for a U$^{4+}$ ion
will always be quadrupolar.  The selection rules for Ce$^{3+}$ ions
restrict the two-channel magnetic Kondo effect to one of three possible doublet
ionic ground states in hexagonal symmetry and
the lone doublet ionic ground state in cubic symmetry.  The dynamical selection
rule apparently excludes two-channel Kondo behavior for
Yb$^{3+}$ ions. \\


{\it Submitted to SCES'92, Sendai Japan, September, 1992}\\
\parindent=1cm

The multi-channel Kondo model was first introduced by Nozi\`{e}res and
Blandin\cite{nozbland}.
The Hamiltonian for the model in which an impurity is placed at the center of a
lattice with $N_s$ sites is
$$H = \sum_{k\sigma\alpha} \epsilon_k c^{\dag}_{k\sigma\alpha}c_{k\sigma\alpha}
- {{\cal J} \over N_s} \vec S_I\cdot
\sum_{k,k',\sigma,\sigma',\alpha} \vec S_{\sigma,\sigma'}
c^{\dag}_{k\sigma\alpha}c_{k\sigma'\alpha} \leqno(1)$$
where $c^{\dag}_{k\sigma\alpha}$ creates a conduction state of wave vector $k$,
spin index $\sigma=\uparrow,\downarrow$, and channel index
$\alpha$ which may take one of $N_{ch}$ values.  The channel index is an extra
internal degree of freedom whose meaning is dependent upon the
particular impurity of interest; it will either represent a local orbital
degree of freedom or the magnetic index of the conduction states in
cases of physical interest.
The conduction states couple to the impurity spin $S_I$ via antiferromagnetic
exchange ${\cal J}<0$. Note
that the spin in each channel couples equivalently to the impurity.
Based upon perturbation theory in $1/N_{ch}$,
Nozi\'{e}res and Blandin argued that so long as $N_{ch}/2>S_I$ this
overcompensated state will be obtained, and
one will also have a non-Fermi liquid excitation spectrum.
Non-perturbative calculations have confirmed this argument in
detail\cite{nrgold,wiegtsv,anddest,sacschlott,confth,nrgnew,aflupaco,boson,nca}.

It has been proposed that this unusual state is realized in exotic Kondo
models.  Specifically
electron assisted atomic tunneling in a double well\cite{vladzow,murgui} and
the quadrupolar Kondo effect\cite{coxold} map onto the
overcompensated  $N_{ch}=2,S_I=1/2$ version of Eq. (1). In this case the
non-Fermi liquid excitation spectrum generates a $T\ln(T_K/T)$
behavior in the specific heat $C(T)$ and leads to a residual entropy per
impurity of $\Delta S(T=0) = (1/2)k_B\ln(2)$.
The discovery of the heavy fermion alloy Y$_{1-x}$U$_x$Pd$_3$ which displays a
$T\ln(T)$ specific heat over one and a half decades of
temperature and possesses clear evidence for a residual entropy of
$(k_B/2)\ln(2)$ per uranium ion provides compelling evidence for the
reality of the quadrupolar Kondo effect\cite{uypd3}.

It is prudent to ask whether the two-channel Kondo model may have broader
applicability to the heavy fermion materials.
The purpose of this article is to answer this question in the affirmative.
I will state five selection rules which contain
the minimal conditions requisite to have the low energy scale physics for a
single ion
described by a two-channel Kondo model.

Fig. 1 displays the basic picture of the states of an impurity
lanthanide/actinide ion  and conduction electrons
which are minimally required to achieve a two-channel Kondo model description
of low energy scale physics.  This restriction to two configurations
is sufficient for U$^{4+}$(5f$^2$) because the two lowest excited
configurations must have odd numbers of electrons and therefore have
at least doubly degenerate crystal field states. For definiteness, we assume
the first excited configuration is $f^1$.
For Ce$^{3+}$, we need to augment this
picture to include three configurations as shown in Fig. 2, because the excited
$f^0$ state must be a singlet.

Regardless, we see
that two configurations of the impurity ion have doublets as the lowest crystal
field states.  These states
span vector spaces which transform under the irreducible representations
$\Gamma_{grd},\Gamma_{ex}$ of the group $\bar G\times {\cal T}$
where $\bar G$ is the double point group of the crystal and ${\cal T}$ is the
group of time reversal (containing the identity and time
reversal operators).  The states in the $\Gamma_{grd,ex}$ spaces have labels
$\alpha_{grd,ex}$.
The extension to include ${\cal T}$ covers groups such as $C_6$ under which
certain pairs of irreducible representations are
complex singlets whose degeneracy is not assured by $\bar G$ but is assured by
${\cal T}$.  The double group is required because one of the
configurations will always contain an odd number of electrons.
The subscripts
$grd,ex$ refer to ground and excited configurations.

We express the conduction operator  $c^{\dag}_{\vec k,\sigma}$ which
creates a Bloch state of momentum $\vec k$, spin $\sigma$, and energy
$\epsilon$ in a symmetry adapted basis around the impurity, i.e.,
$$c^{\dag}_{\vec k,\sigma} = \sum_{\Gamma_c,\alpha_c}
a_{\Gamma_c,\alpha_c}(\epsilon)c^{\dag}_{\epsilon\Gamma_c\alpha_c} \leqno(2)$$
where $\Gamma_c,\alpha_c$ label irreducible representations of $\bar G\times
{\cal T}$ where the point group $\bar G$
is defined at the impurity site.
The local conduction states are
derived from partial waves in a plane-wave basis, or from suitable linear
combinations of ligand orbitals in a tight-binding basis.

The Anderson Hamiltonian for U$^{4+}$ ions then takes the form
$$H = H_{cond} + H_{grd} + H_{ex} + H_{mix} \leqno(3)$$
with
$$H_{cond} = \sum_{\epsilon\Gamma_c\alpha_c}\epsilon
c^{\dag}_{\epsilon\Gamma_c\alpha_c}c_{\epsilon\Gamma_c\alpha_c} \leqno(4)$$
$$H_{grd} = E(f^2) \sum_{\alpha_{grd}}
|\Gamma_{grd}\alpha_{grd}><\Gamma_{grd}\alpha_{grd}| \leqno(5)$$
$$H_{ex} =  E(f^1) \sum_{\alpha_{ex}}
|\Gamma_{ex}\alpha_{ex}><\Gamma_{ex}\alpha_{ex}| \leqno(6)$$
and
$$H_{hyb} = {1\over
\sqrt{N_s}}\sum_{\epsilon}\sum_{\Gamma_c\alpha_c}\sum_{\alpha_{grd},\alpha_{ex}}
V[\Gamma_c\alpha_c;\alpha_{grd};\alpha_{ex}]
[|\Gamma_{grd}\alpha_{grd}><\Gamma_{ex}\alpha_{ex}|c_{\epsilon\Gamma_c\alpha_c}
+h.c.]
\leqno(7)$$
where $V[\Gamma_c\alpha_c;\alpha_{grd};\alpha_{ex}]$ includes the single
particle matrix element, a reduced matrix element expressing the
attachment probability for adding an $f$ electron to get this $f^2$ state from
this $f^1$ state, and a Clebsch-Gordan coefficient in the
crystal field representation basis.
Again, the restriction to $f^1$ for the excited configuration is purely  a
matter of convenience for the exposition purposes here.
For the Ce$^{3+}$ ions, we interchange ground and excited levels and (with
$E(f^0)$ set at zero in this case) add the hybridization
term
$${V_0 \over \sqrt{N_s}}\sum_{\epsilon\alpha_{grd}}
[|\Gamma_{grd}\alpha_{grd}><f^0| c_{\epsilon\Gamma_{grd}\alpha_{grd}} +
h.c.]~~.\leqno(8)$$

In addition to discussing the symmetry properties of the states themselves, it
is important to discuss the symmetry properties of the ground
configuration tensor operators which
live in the product space transforming according to
$\Gamma_{grd}^{ket}\otimes\Gamma_{grd}^{bra}$. The superscripts are a reminder
that the tensors are formed from outer products of the states.  The form of the
low energy scale interactions which will correspond to the
two-channel coupling of Eq. (1) are entirely specified by the symmetry
properties of these tensors.  The interactions arise when we integrate out
the virtual charge fluctuations to the excited configuration to derive an
effective interaction between the conduction electrons and the
ground configuration degrees of freedom of magnitude $\sim V^2/\Delta E$, where
$\Delta E$ is the interconfiguration energy
difference\cite{schwolf}.  [Note:  all notation for point group
representations used in
this paper follow those of Koster {\it et al.}\cite{koster}.]

We now state the selection rules which will minimally ensure that the effective
Hamiltonian
at low energy scales derived from an underlying Anderson Hamiltonian has the
two-channel $S=1/2$ Kondo form:\\
\begin{quote}{\it Selection Rule 1 (Ground Doublet Selection Rule)}:  Under the
action of the crystal field, the
lowest state of the lowest angular momentum multiplet of the ground
configuration
should be a degenerate doublet which transforms as the irreducible
representation
$\Gamma_{grd}$ of the group $\bar G\times {\cal T}$.
\end{quote}
\begin{quote}{\it Selection Rule 2 (Excited Doublet Selection Rule)}: Under the
action of the crystal field, the lowest state of the lowest
lying angular momentum of the excited configuration must be a degenerate
doublet transforming as a representation $\Gamma_{ex}$ of the group
$\bar G\times{\cal T}$.
\end{quote}
\begin{quote} {\it Selection Rule 3 (Hybridization Selection Rule):} The
conduction band must contain states which, when projected to the
impurity site, transform as the direct product representation $\Gamma_c$ such
that
$\tilde\Gamma_c=\Gamma_{grd}\otimes\Gamma_{ex}$.
\end{quote}
\begin{quote} {\it Selection Rule 4 (Tensor Selection Rule):}  When $\Gamma_c$
is a reducible representation of the group $\bar G\times {\cal
T}$ of the form
$\Gamma_c=\Gamma_{c1}\oplus\Gamma_{c2}$ where $\Gamma_{c1,2}$ are irreducible
doublet representations of $\bar G\times{\cal T}$, then
the tensors $\Gamma_{grd}^{ket}\otimes\Gamma_{grd}^{bra}$ which are off
diagonal
in the $\Gamma_{grd}$ space must be contained in the space of the product
representation
$\Gamma_{c1}\otimes\Gamma_{c2}$ and not in
$(\Gamma_{c1}\otimes\Gamma_{c1})\oplus(\Gamma_{c2}\otimes\Gamma_{c2})$.
\end{quote}
\begin{quote}{\it Selection Rule 5 (Dynamic selection rule):} When
$\Gamma_{grd}$ is the lowest level of an odd number electron configuration,
the exchange coupling generated by virtual
excitations to $\Gamma_{ex}$ must be larger than the coupling induced by
virtual charge fluctuations to excited singlet states.\end{quote}

Selection Rule 1 ensures that the lowest impurity states have internal degrees
of freedom so that
a Kondo effect is possible.
Selection Rules 2,3 are necessary for two-channel behavior in conjunction with
Rule 1.  The excited doublet state labels {\it are}
the channel indices, and if
the conduction band doesn't have local
$\Gamma_c=\Gamma_{grd}\otimes\Gamma_{ex}$ symmetry states, no Kondo effect is
possible.

The basis for Selection Rule 4 is an examination of the tensor operator
structure.  This rule is irrelevant for cubic structure because the
irreducible
$\Gamma_8$ representation is the only quartet of conduction states allowed for
the $\Gamma_3$ and $\Gamma_{6,7}$ doublets of the different
configurations.  It is essential for the lower symmetry crystal syngonies.  To
see why, note that we
form the exchange term of Eq. (1) by coupling tensors
of the impurity states $\Gamma_{grd}$ to those with the same symmetry derived
from
the conduction states $\tilde\Gamma_c$.  In the lower symmetry syngonies of
interest (hexagonal, tetragonal, rhombohedral) $\tilde\Gamma_c$ is
reducible, decomposing into $\Gamma_{c1}\oplus\Gamma_{c2}$.  The magnitude and
antiferromagnetic sign are set by the integration
out of virtual fluctuations to the $\Gamma_{ex}$ states; hence, as in the
conventional Anderson impurity, the exchange ${\cal J}<0$ always.
Now, the tensor operators forming the basis for
$\Gamma_{grd}^{ket}\otimes\Gamma_{grd}^{bra}$ include two which are diagonal in
the
$\Gamma_{grd}$ indices and two which are
off-diagonal corresponding to $S_I^{(1,2)}$ in the pseudo-spin 1/2 space.
The identity operator in the $\Gamma_{grd}$ space gives charge scattering,
while the other diagonal term
corresponds to the $S^{(3)}_I$ component of the impurity
pseudo-spin.  This is always contained in both the tensor spaces
$\Gamma_{c1}^{ket}\otimes\Gamma_{c1}^{bra}$
and $\Gamma_{c2}^{ket}\otimes\Gamma_{c2}^{bra}$.   If the off-diagonal impurity
operators were contained in these tensor product spaces, then
no symmetry conditions would ensure the exact equality of exchange coupling
constants between the channels (now indexed simply by the irreducible
representation labels $\Gamma_{c1,2}$). This is {\it always} the case in
rhombohedral
symmetry, so that two-channel coupling will not be generically present in this
case for $f$-ions.  However, if the off-diagonal operators are
contained in the mixed-direct product
$\Gamma_{c1}^{ket}\otimes\Gamma_{c2}^{bra}$, then the ``spin-flip'' conduction
tensors must mix states
of the two doublets, and the channel degeneracy is automatically ensured.

The Table specificies all the possible two-channel $S=1/2$ combinations of
$\Gamma_{grd},\Gamma_{ex},\Gamma_c$ states
for U$^{4+}$ and Ce$^{3+}$ ions\cite{butsplitdoub}.

Selection Rule 5 follows from scaling along the lines of Nozi\`{e}res and
Blandin\cite{nozbland}, and from
DNCA analysis for the model Ce$^{3+}$ ion\cite{nca}. Specifically, we define
two crossover temperatures $T^{(I,II)}_x$ where the
superscript refers to single or two-channel crossover.  Consider the case in
cubic symmetry, with $\Gamma_{grd}=\Gamma_7$,
$\Gamma_{ex}=\Gamma_3$, and $\Gamma_c=\Gamma_8$.  Define dimensionless
effective exchange coupling constants $\tilde g_7,\tilde g_8$ by
$$\tilde g_7 = {N(0)V^2_7 \over E_0 - E(f^0)} ~~~;~~~ \tilde g_8 =
{N(0)V_8^2\over E_0 - E(f^2)} \leqno(9)$$
where $E_0$ is the ground state energy, $N(0)$ is the conduction electron
density of states,
and $V_{7,8}$ the hybridization matrix elements with $\Gamma_{7,8}$ partial
waves.  The Kondo scale for
the single channel model is $T_0^{(I)}\simeq D\exp(1/2\tilde g_7)$, and for the
two-channel model is $T_0^{(II)}\simeq D\exp(1/2\tilde g_8)$
to leading exponential accuracy.  Then for $T$ below the crossover scale
$T_x^{(I)}$
$$T^{(I)}_x \approx {T_0^{(I)}\over 3}({1\over \tilde g_8} - {1\over \tilde
g_7})^{3/2} \leqno(10)$$
single channel behavior will dominate for $|\tilde g_7|>|\tilde g_8|$.  For $T$
below the crossover scale $T_x^{(II)}$ given by
$$T^{(II)}_x \approx {T_0^{(II)}\over 4}({1\over \tilde g_7} - {1\over \tilde
g_8})^2 \leqno(11)$$
two-channel physics will dominate for $|\tilde g_8|>|\tilde g_7|$.
Practically, this is tested by examining the sign of the thermopower,
given the particle-hole asymmetry of the model.
Dominant $f^0$-induced one-channel coupling will tend to produce positive
thermopower, while dominant $f^2$-induced two-channel coupling will
tend to produce a negative thermopower.  For hexagonal symmetry, consider $D_6$
for concreteness.  Analogous arguments go through provided
$\Gamma_{grd}=\Gamma_9 \sim |\pm3/2>$, and $\Gamma_{ex}=\Gamma_{5,6}$, with
$\Gamma_c=\Gamma_7\oplus\Gamma_8$.  For Yb$^{3+}$ ions with
$\Gamma_{6,7}$ ground states in cubic symmetry or $\Gamma_9$ in hexagonal
symmetry similar arguments go through, with $f^n\to f^{14-n}$. However,
the very large $f^{12}-f^{13}$ splitting (order 10 eV) makes it unlikely that
the conditions of Eq. (11) can be realized.  Note that all
treatments with infinite Coulomb repulsion miss this physics.

U$^{4+}$ ions with $f^2$ ground configurations have Hund's rules angular
momentum $J=4$, so that non-Kramers doublets are possible.  Consider
the
ground doublets for the hexagonal and tetragonal syngonies.
All of these doublets have the
property that the non-trivial diagonal operator transforms like the
$z$-component of a real spin, while the off-diagonal elements are
quadrupolar and contained only in direct products of two distinct irreducible
representations of local
conduction states.  The physical reason is simple, and easily understood by
considering states with $M_J=\pm 1$ pair of states in the presence of a
crystal field Hamiltonian of pure axial character, viz $H_{cef} = \Delta_{cef}
[3J^2_z-J(J+1)]$.  This term is even under time reversal and
thus maintains doublet degeneracy of the $\pm 1$ states.  Hence, the
non-trivial diagonal tensor is just $|1><1|-|-1><-1|$ which transforms
like $J_z$ as restricted to the doublet.  The off diagonal tensor must change
the angular momentum by two units, and hence must have
quadrupolar character.  Turning now to $\Gamma_{ex},\Gamma_{c1,2}$ in the same
axial field, we must have doublets of the form $\pm (2n+1)/2$, since the
excited configuration has an odd number of electrons and the conduction states
always transform as double valued representations which are
descended from half integral angular momentum in the full isotropic symmetry.
Hence, the off diagonal conduction tensors in a given
representation can only change angular momentum by an {\it odd} number of
units, and cannot `flip' the  impurity spin.  However, it is
possible to form tensors from the cross products
$\Gamma_{c1}\otimes\Gamma_{c2}$ which can change the angular momentum index by
two units.
For example, for conduction states derived from $j=5/2$ partial waves, the
operator $|3/2><-1/2|$ changes the angular momentum by two units.
Note that the results we have discussed are properties only of the
representations,
but easily illustrated in this pure axial limit.

{}From the discussion of the preceding paragraph
it is apparent that any Kondo model derived from a degenerate doublet ground
state of U$^{4+}$ ions in hexagonal or
tetragonal symmetry will be of the quadrupolar form, because the only
degenerate levels in ground or excited states are doublets, time
reversal guaranteeing the channel degeneracy (indexed by the excited state in
effect).  To make the idea more explicit, assume for
definiteness that we have hexagonal $D_6$ point symmetry and
$\Gamma_{grd}=\Gamma_5$, $\Gamma_{ex}=\Gamma_7$ in an excited $f^1$
configuration.  This
yields $\tilde\Gamma_c=\Gamma_5\otimes\Gamma_7=\Gamma_7\oplus\Gamma_9$.  Taking
the simple axial crystal
field model and using conduction plane waves in a $j=5/2$ partial wave
manifold,
representative states in $J,M_J$ form are $|\Gamma_5\pm>=|4,\pm>$,
$|\Gamma_7\pm> = |5/2,\pm1/2>$, and $|\Gamma_9\pm>=|5/2,\pm
3/2>$.
Let us reorganize the labelling of the conduction states.  Define channel 1 as
labelling the states created by the pair of operators
$c^{\dag}_{k\Gamma_7+},c^{\dag}_{k\Gamma_9-}$, and channel 2 as labelling the
pair of states
created by $c^{\dag}_{k\Gamma_9+},c^{\dag}_{k\Gamma_7-}$.  Now let $\alpha$ be
the spin index, equal to $\pm$, and $\mu=1,2$ be the channel
index.  Denote channel spin operators by $\tau^{(i)},i=1,2,3$.   Thus, for
example,
$c^{\dag}_{k,+,1} = c^{\dag}_{\Gamma_7+}$.
By performing a Schrieffer-Wolff transformation, with the interconfiguration
energy splitting given by
$\epsilon_f=E(f^2\Gamma_5)-E(f^1\Gamma_7)$, we obtain the Kondo coupling
$$H_{Kondo} = -{1\over N_s}\sum_{i,k,k',\alpha,\alpha',\mu}{\cal
J}^{(i)}S^{(i)}_I S^{(i)}_{\alpha,\alpha'}
c^{\dag}_{k\alpha\mu}c_{k'\alpha'\mu} - {{\cal K}\over N_s} S^{(3)}_I
\sum_{k,k',\mu,\alpha} \tau^{(3)}_{\mu,\mu}c^{\dag}_{k\alpha\mu}c_{k'\alpha\mu}
\leqno(13)$$
where ${\cal J}^{(1,2)}= V_7V_9/\epsilon_f$, ${\cal
J}^{(3)}=(V_7^2+V_9^2)/2\epsilon_f$, and ${\cal K}=(V_7^2-V_9^2)/2\epsilon_f$.
Here
$V_{7,9}$ are the hybridization matrix elements coupling the $\Gamma_{7,9}$
conductions states to the impurity. Note that:
i) this exchange Hamiltonian is intrinsically anisotropic but the diagonal
(${\cal J}^{(3)}$) term is antiferromagnetic which is sufficient to
ensure the Kondo effect, and (ii) this Hamiltonian has the peculiar term
coupling diagonal spin and channel spin operators.  These are not of
concern, since it is now well established that exchange anisotropy is
irrelevant in the $N_{ch}=2,S_I=1/2$ model\cite{aflupaco},
and scaling calculations about
the non-trivial fixed point indicate that the spin-channel spin coupling is
marginally irrelevant\cite{pangpriv}.

Hence, we have demonstrated that the mapping of low energy scale properties to
the
two-channel quadrupolar Kondo model is  robust for U$^{4+}$ ions with doublet
ground states in hexagonal and tetragonal symmetries in that {\it all} such
doublets will be described by this model on coupling to the
conduction states.  We have also shown that under more restrictive conditions
the model will apply to Ce$^{3+}$ ions in cubic and hexagonal
symmetry, but is unlikely to apply to Yb$^{3+}$ ions.

{\it Acknowledgements}.
I would like to thank I. Affleck, J.W. Allen, T.-L. Ho, M.B. Maple,
A.W.W. Ludwig, V. Mineev, H.-B. Pang,
C. Seaman,
and J.W. Wilkins for useful conversations.
This research was supported by the U.S. Department of Energy,
Office of Basic Energy Sciences, Division of Materials Research.

\pagebreak

\pagebreak

{\bf Table}.  Ground, excited, and conduction states for proposed two-channel
Kondo models for \ufp,\ctp ions.  \\
Notation follows Koster {\it et al.}\cite{koster}.

\vspace{.2in}

\begin{tabular}{|c|c|c|c|c|c|}\hline
Ion & Ground & Point & $\Gamma_{grd}$ & $\Gamma_{ex}$&  $\Gamma_c$ \\
& Config.&   Group &&&   \\\hline\hline
 U$^{4+}$&  $5f^2(J=4)$&  Cubic($O$)&  $\gth (E)$ & $\gse$&
 $\gei=\gth\otimes\gse$\\\hline
 \ufp&  \ucon & Hexag.($D_6$) & $\gfi (E_1)$&  $\gse $ & $\gse\oplus\gni$ \\
&&&& $\gei$ & $\gei\oplus\gni$\\
&&&& $\gni$ & $\gse\oplus\gei$\\
&&& $\gsi (E_2)$ & $\gse$ & $\gei\oplus\gni$\\
&&&& $\gei$ & $\gse\oplus\gni$ \\
&&&& $\gni$ & $\gse\oplus\gei$ \\\hline
  \ufp & \ucon  &Tetrag.($D_4$)&  $\gfi (E)$ & $\gsi$ or $\gse$&
 $\gse\oplus\gsi$\\\hline
 \ctp& \ccon & Cubic($O$)&  $\gse$ & $\gth$ & $\gei$ \\\hline
 \ctp & \ccon & Hexag($D_6$) & $\gni$ & $\gfi$ or $\gsi$&
 $\gsi\oplus\gse$\\\hline
 \end{tabular}

 \pagebreak

 {\bf Figure Captions}\\

 {\bf Figure 1}.  Energy level scheme for \ufp ion.  Energy runs vertically.
 Only two configurations are kept for simplicity; so long as the $f^3$ has a
doublet level lowest
 the same physics will ensue. We show only the lowest states.  \\

 {\bf Figure 2}.  Level scheme for \ctp ions.  Because the $f^0$ is a singlet,
it must be retained, and the singlet channel and two-channel
 Kondo effects compete.  \\


\begin{thebibliography}{99}
\bibitem{butsplitdoub} A split doublet could display two-channel behavior above
a crossover scale below $T_0$.
\end{thebibliography}
\end{document}